# Ultrafast Drop Movements Arising from Curvature Gradient

Cunjing Lv[1,2,†], Chao Chen[1,2,†], Yin-Chuan Chuang[3], Fan-Gang Tseng[3,4], Yajun Yin[1],

Quanshui Zheng[1,2,*]

[1] Department of Engineering Mechanics, Tsinghua University, Beijing 100084, China

[2] Center for Nano and Micro Mechanics, Tsinghua University, Beijing 100084, China

[3] Department of Engineering and System Science, National Tsing Hua University, Hsinchu 30013, Taiwan

[4] Research Center for Applied Sciences, Academia Sinica, Taipei 11529, Taiwan

**We report experimental observation of a kind of fast spontaneous movements of water drops on surfaces of cones with diameters from 0.1 to 1.5 mm. The observed maximum speed (0.22 m/s) under ambient conditions were at least two orders of magnitude higher than that resulting from any known single spontaneous movement mechanism, for example, Marangoni effect due to gradient of surface tension[1-4]. We trapped even higher spontaneous movement speeds (up to 125 m/s) in virtual experiments for drops on nanoscale cones by using molecular dynamics simulations. The underlying mechanism is found to be universally effective - drops on any surface either hydrophilic or hydrophobic with varying mean curvature are subject to driving forces toward the gradient direction of the mean curvature. The larger the mean curvature of the surface and the lower the contact angle of the liquid are, the stronger the driving force will be. This discovery can lead to more effective techniques for transporting droplets.**

Self-propelled or spontaneous movements of liquid drops on solid surfaces constitute an important class of surface phenomena[4,]. These types of drop transportation are of value in many industrial applications, such as lab-on-a-chip technology[5,6], drug screening, and integrated DNA analysis devices[7].

Most observed spontaneous movement phenomena could be understood as caused by surface energy gradient, i.e. representations of Marangoni effect[1-4]. Many techniques, such as thermal[2], chemical[8], electrochemical[9], and photochemical methods[10,11], can be used to make a flat surface possess continuously varying surface energy or liquid contact angle. Drops on such surfaces tend to move along the gradient direction of contact angle toward the region with lower contact angle or

---

surface energy. Although a spontaneous movement phenomenon has also been observed for drops on substrates with uniform surface energy but varying surface roughness[12], its mechanism can still be categorized as a generalized Marangoni effect because of the varying effective contact angle. The spontaneous moving speeds resulting from the single contact angle gradient mechanism typically range from micrometers per second to millimeters per second[13], which seem, however, too slow for practical applications.

The main obstacle to drop movement on a solid surface arises from contact angle hysteresis. In order to overcome this difficulty, energy feeding or other external factors have been introduced. For example, vibrating or heating a surface with gradient of surface energy or roughness can result in faster drop movements[13-15]. Fast radial spontaneous moving speeds up to 0.3 m/s were observed[13] for condensation drops coming from saturated steam (100°C) on a silicon wafer with a radial gradient of surface energy. The up-to-date record of the observed spontaneous moving speed at ambient conditions is about 0.5 m/s[16,17], that has resulted from a chemically patterned nanotextured surface combined with wedge-shaped gradient.

Here we report a mechanism that can result in ultrafast spontaneous movements of drops, with speeds over 100 m/s to be expected. In contrast with Marangoni effect, neither surface energy gradient nor surface roughness variation but varying surface curvature is involved in this mechanism.

At first, we describe our experimental observation of this kind of fast spontaneous movements for water drops on small cones. Three cone samples (Cones I, II, III) were fabricated from glass tubes by using the electrostatic spinning method, each with a radius-of-curvatures range of 0.1-1.0 mm, 0.3-1.0 mm, 0.3-1.5 mm and an axial lengths of 5.6mm, 4.8mm, 9.1mm, respectively.

The video frames shown in Figure 1(a) illustrate a typical spontaneous movement of a 1 μL water drop under ambient conditions on Cone I with a plain glass surface (water contact angle $\theta \approx 28°$). With the drop moving from the cone tip toward the direction of increasing curvature radius, evolution of its spontaneous movement velocity is measured and plotted as the green triangles in Figure 1(b). The blue circles and red squares in Figure 1(b) show typical velocity evolutions of 1 μL water drops on the same Cone I but with its surface $O_2$ plasma treated (contact angle $\theta \approx 5°$) and MTS nanotextured plus $O_2$ plasma treated (contact angle $\theta \approx 0°$), respectively. These results indicate that the lower the water contact angle is, the faster the spontaneous movement can become.

Furthermore, similar observations were made for 1μL water drops on the two larger cones (II and III) with differently treated surfaces. The plots in Figure 1(c) show typical velocity evolutions on Cones I, II, III treated by $O_2$ plasma ($\theta \approx 5^0$), indicating the fastest spontaneous movement on the smallest cone. The observed maximal spontaneous movement speed (~0.22 m/s, see Fig. 1(b)) is amazingly two orders in magnitude higher than those by any other known single spontaneous movement mechanism[8,18]. This observation suggests that drops on nanoscale cones may reach ultrafast spontaneous movements.

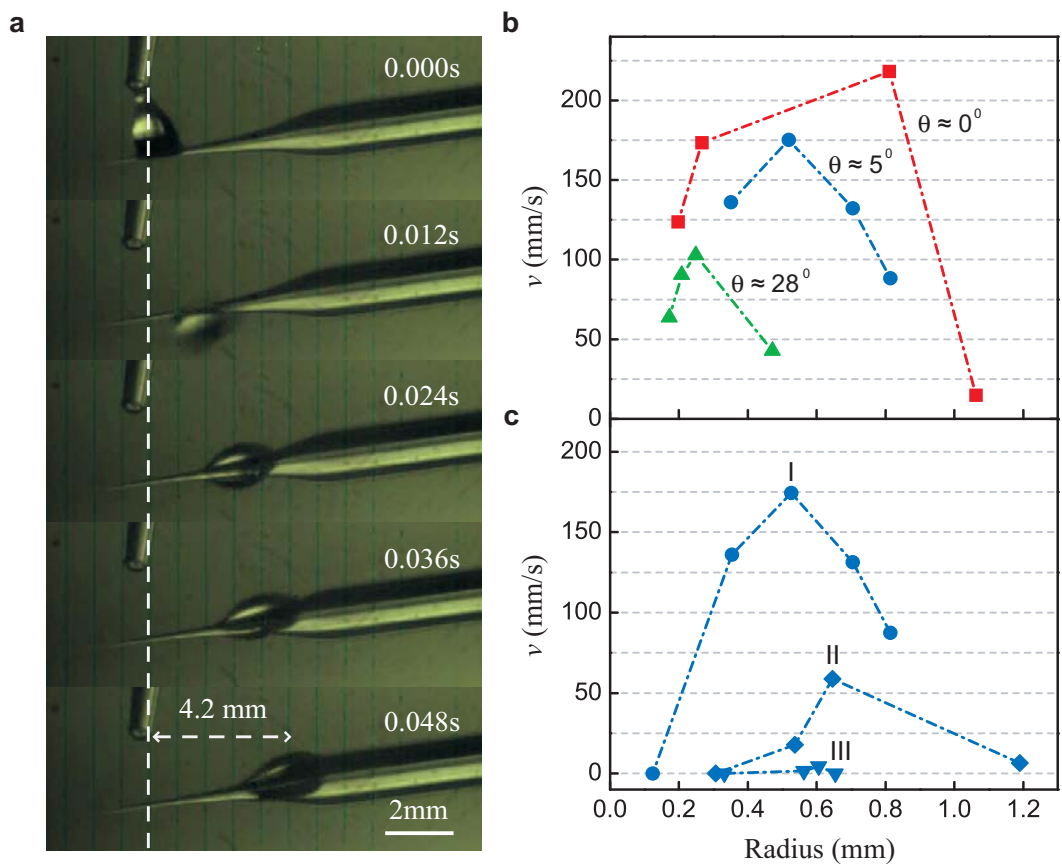

**FIG. 1**. Experimentally observed fast spontaneous movements of 1μL water drops on cones with different radius and surface conditions. (**a**) Video frames of a drop moving on a glass cone with radius range 0.1 - 1.0 mm and plain glass surface (contact angle $\theta \approx 28°$). (**b**) Velocity evolutions of drops moving on the same cone with three different surface conditions (plain, $\theta \approx 28°$; $O_2$ plasma treated, $\theta \approx 5°$; and MTS nanotextured plus $O_2$ plasma treated, $\theta \approx 0°$, see the videos in the Supplementary Information for more details.). (**c**) Velocity evolutions of drops moving on $O_2$ plasma treated cones ($\theta \approx 28°$) with three different radius ranges 0.1-1.0mm (I), 0.3-1.0mm (II), and 0.3-1.5mm (III).

Similar experiments on nanoscale, however, face severe technical challenges. . Instead, we performed virtual tests by using molecular dynamics (MD) simulations based on the platform of LAMMPS[19]. The tiny cone was modeled as a rigid framework of mono-layered atoms that likes a conically rolled graphene, as shown by the inserts in Figure 2(a) and 2(b). The water was modeled by a widely used code (SPC/E[20]) with the same parameters as given in Ref. 21. Lennard-Jones potential, $\phi(r) = 4\varepsilon[(\sigma/r)^{12}-(\sigma/r)^{6}]$, was used to characterize the cone-water van der Waals interaction[22], where $r$ denotes the distance between atoms. By fixing equilibrium distance $\sigma$ at 0.319 nm and adopting values 5.85, 1.95, and 1.50 meV for well depth $\varepsilon$, we yielded three different water contact angles $\theta$ = 50.7° (hydrophilic), 138° (hydrophobic), and nearly 180° (superhydrophobic), respectively. The model cone has a half-apex angle $\alpha$ = 19.5° and a height of 7 nm to its tip. As illustrated in Figures 2(a) and 2(b), we then cut off the tip at the heights of 1.5nm and 3.5nm, respectively, for modeling the movement of a 2nm-diameter water drop (containing 339 atoms) on the outer and inner conical surfaces. In the simulations, the temperature of water was kept at 300 K with Nosé/Hoover thermostat, and the whole system was located in a finite vacuum box. At beginning, the mass center of the drop was fixed at the starting point for 100 ps to reach thermal equilibrium. Then the drop was released to move freely along the conical surface for the next 300 ps.

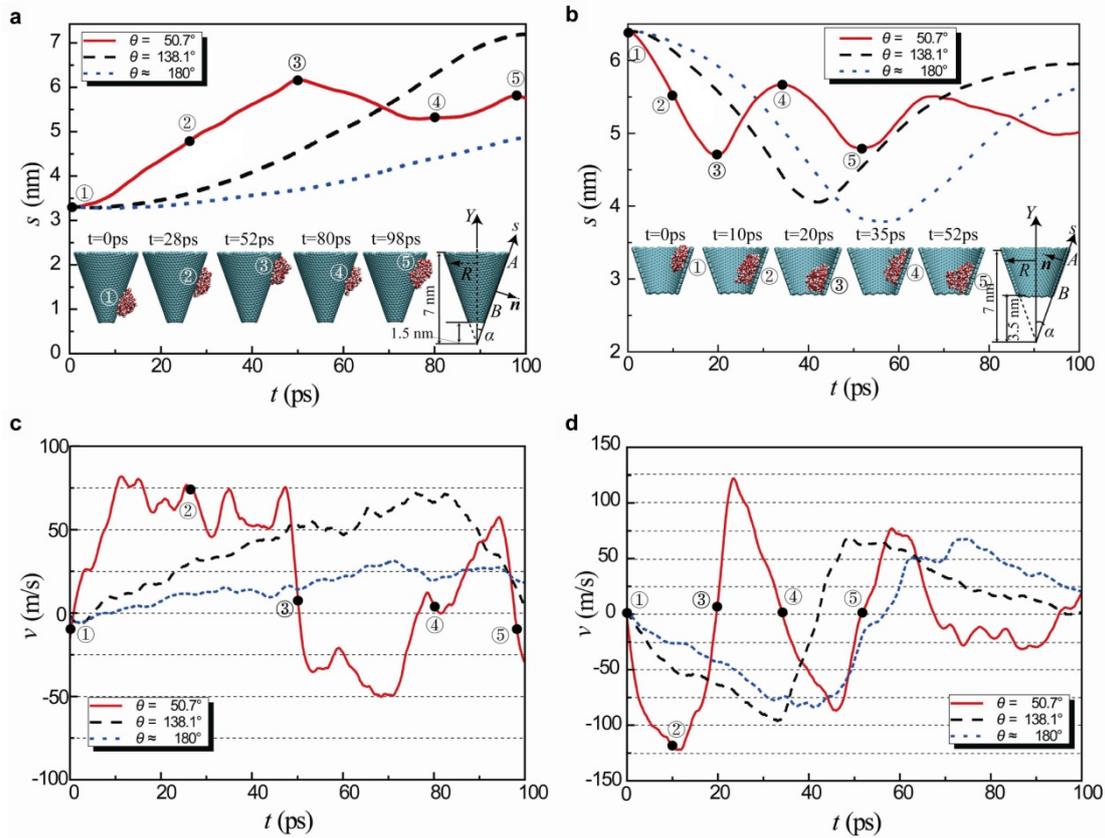

**FIG. 2.** Molecular-dynamics simulated spontaneous moving and bouncing behaviors of a drop consisting of 339 water molecules on the outer surface (**a**,**c**) and inner surface (**b**,**d**) of a cone. The red solid, black dashed and blue dotted trajectories correspond to water contact angles of 50.7°, 138° and near 180°, respectively. Inserts in (**a**) and (**b**) show drop positions on the outer and inner surfaces of the cone respectively at five different moments selected from the MD simulation movie (see Supplementary Information). (**c**) and (**d**) give velocity evolutions corresponding to (**a**) and (**b**).

The MD simulations result in trajectories of the mass center along the meridian line (coordinate *s*), as shown in Figure 2. When the drop was released near the smaller end of the outer surface, it immediately started a spontaneous movement toward the larger end. When approaching the larger end, the drop was starting to bounce back. The red solid curve in Fig. 2(a) depicts the dependence of moving distance *s* of the drop versus time *t* for contact angle $\theta$ = 50.8°, and the inserts show five frames selected from the movie record (see Supplementary Information). Similar spontaneous movement and bouncing were found for the hydrophobic ($\theta$ = 138°) and superhydrophobic ($\theta \approx$ 180°) surfaces, as shown by the black dashed- and blue dotted-lines in Figure 2(a). On the other hand, when the water droplet was released on the inner surface we observed diametrically opposite results, as illustrated in Figure 2(b). The droplet started spontaneous movement toward the smaller end and bounced back at the larger end. The velocity trajectories versus time shown in Figure 2(c) and 2(d) indicate that ultrafast spontaneous movements are possible for drops on cones on nanoscale. The maximum speeds attained on the outer and inner conical surfaces are about 75 m/s and 125 m/s, respectively, and the accelerations increase as the contact angle decreases.

Of particularly interest are that the spontaneous movement happens for drops on not only hydrophilic, but also hydrophobic conical surfaces, and that the moving speeds decrease as increasing the contact angle. To our best knowledge, this is the first report about spontaneous movement of drops on hydrophobic surfaces. Spontaneous movements of drops on conical surfaces have been observed previously, but limited to hydrophilic surfaces[18,23,24]. The reported moving speeds were a few millimeters per second or lower.

To discover the underlying mechanism, we study evolution of the total free energy of the system when quasi-statically moving a liquid drop on a solid surface with a varying curvature. For small drops, we can ignore the gravity potential. Thus, the total free energy can be quantified as[4,25]: $U = \gamma_{LV}(A_{LV} - A_{SL}\cos\theta)$, where $A_{LV}$ and $A_{SL}$ denote the liquid-vapor and solid-liquid interfacial areas, $\gamma_{LV}$ is the liquid-vapor interfacial energy, and $\theta$ is the liquid contact angle. We used a finite element code, Surface Evolver[26], to determine the precise shapes of water drops that are quasi-statically staying on the outer and inner surfaces of cones with different contact angles. For each specific location of the drop the total free energy $U$ was minimized as a function of $A_{LV}$ and $A_{SL}$ for the fixed drop volume. The plots in Figure 3(a) show the resulting free energies for a 1 μL water drop on the outer surface (solid line) and inner surface (dashed line) of a cone versus the core radius where the drop located. Three different contact angles 120°, 90°, and 75° were considered, showing a similar tendency of the free energy $U$. The results clearly show that the driving force is always pointing to the cone diameter increasing or decreasing direction whenever the drop is locating on the outer or inner surface, respectively. Again, this directionally moving property exists regardless of the surface being hydrophilic or hydrophobic.

We can further draw a unified picture instead of the different descriptions on outer and inner conical surfaces. It is known from the theory of differential geometry[27] that the shape of any surface is determined by its two principal curvatures $\kappa_1$ and $\kappa_2$, or equivalently, its mean curvature $H = (\kappa_1 + \kappa_2)/2$ and Gaussian curvature $K = \kappa_1 \kappa_2$, and that the principal curvatures change their signs if we invert the surface outer normal. Hereafter, we always appoint the normal of a surface as directing toward the drop. Thus, the principal curvatures of a cone for a drop located on its outer and inner surfaces at distance $s$ from its tip are $\kappa_1 = \mp 1/(s \tan \alpha)$ and $\kappa_2 = 0$. Using $H$ instead of the cross-section diameter $D$, we can redraw the Figure 3(a) into Figure 3(b). From the latter we have three major observations: (i) $U$ is a smooth function of $H$ with monotonously decreasing slope so that the driving force, $F$, is always pointing in the $H$-increasing direction, (ii) function $U$ is independent of half-apex angles ($\alpha = 19.5°$ and $30^0$), and (iii) the lower the surface contact angle is, the larger the driving force can become.

The plots in Figure 3(c) show respective evolutions of the liquid-vapor interfacial area ($A_{LV}$) and the solid-liquid interfacial area ($A_{SL}$) versus $H$, from which it is not difficult to see that $dA_{LV}/dH < 0$ and $dA_{SL}/dH > 0$. The meaning of the relations will be discussed later.

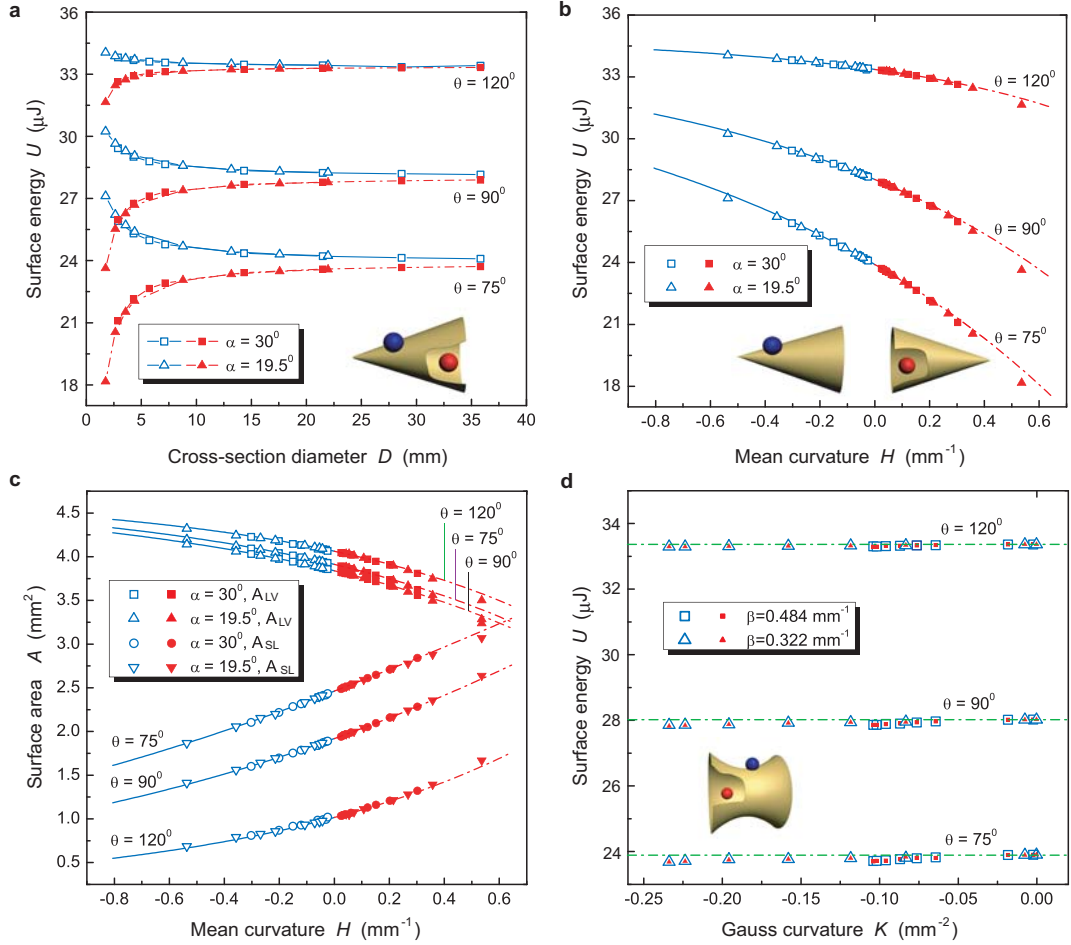

**FIG. 3.** Numerical results of total free energy $U$ and interfacial areas $A_{SL}$ and $A_{LV}$ of a 1 $\mu$L water drop on curved solid surfaces with water contact angle $\theta$ = 120°, 90°, and 75°. (**a**) $U$ versus cross-section diameter $D$ of a cone for the drop on the outer (hollow dots) and inner (solid dots) surfaces. The square and triangle dots represent halp-apex angles of the cone $\alpha$ = 30º and 19.5º, respectively. (**b,c**) $U$, $A_{SL}$, and $A_{LV}$ as functions of mean curvature $H$. The hollow blue and solid red dots represent the drop on inner and outer conical surfaces, and the solid blue and dash-dotted red lines represent the drop on outer and inner surfaces of a sphere. (**d**) $U$ versus Gauss' curvature $K$ for the drop on outer (hollow blue dots) and inner (solid red dots) catenoid surfaces. The square and triangle dots represent $\beta$=0.484mm$^{-1}$ and $\beta$=0.322mm$^{-1}$, and the horizontal dash-dotted lines are results for the drop on flat surfaces.

How does Gauss curvature ($K$) affect the spontaneous movement? Since the Gauss curvature of cones is always equal to zero, we have to study other types of shape. Here we particularly study the free energy of a 1μL water drop on a class of catenoidal surfaces that can be described by $\cosh(\beta z) = \beta\sqrt{x^2 + y^2}$, where $\beta$>0 is the shape parameter and {$x,y,z$} is a cartesian coordinate system. The mean curvature $H$ of each such surface is identically equal to zero while its Gauss curvature, $K = -\beta^2 \cosh^{-4}(\beta z)$, is always negative. We simulated the shapes of the drop on different positions of the catenoidal surfaces by using the same technique as for cones.

The results shown in Figure 3(d) demonstrate independence of the free energy from the Gauss curvature. We guess that this independence would be generally held for any shapes. To check this conjecture, we also plot in Figures 3(b) and 3(c) the free energy $U$ and interfacial areas $A_{LV}$ and $A_{SL}$ of a 1 μL water drop on the outer (solid blue lines) and inner (dash-dotted red lines) spherical surfaces of radius $R$[28]. Although the Gauss' curvature $K = 1/R^2$ for spherical surfaces are nonzero while that for conical surfaces are always equal to zero, one can see from Figures 3(b) and 3(c) that $U$, $A_{LV}$, and $A_{SL}$ do not change with $K$. Our later result, Eq. (2) will further support this conjecture.

Based on a dimensional analysis, we can directly generalize the results given in Figure 3 for liquid drops with different volumes (marked with an asterisk). If $p > 0$ is a proportional coefficient defining the ratio of the drop size, then parameters for the proportioned drop will be $V^* = p^3 V$, $D^* = pD$, $A^* = p^2 A$, $U^* = p^2 (\gamma_{LV}^*/\gamma_{LV})U$, $H^* = H/p$, and $K^* = K/p^2$, where $V$ is the volume (1 μL) of the initially examined water drop. Thus, the same tendencies as shown in Figure 3 can also be seen for a generic drop.

It is known from a recent study using MD simulations that the water contact angles $\theta$ on inner or outer surfaces of single-walled carbon nanotubes with various diameters decrease remarkably with increasing of the tube mean curvature $H$[29]. Therefore, not only interfacial areas $A_{SL}$ and $A_{LV}$, but also contact angle $\theta$ of a liquid drop on a conical surface are functions of $H$. Consequently, the driving force ($F$) that results in spontaneous movements of a drop on a conical surface can be expressed as

$$F = \mp \frac{dU}{ds} = -\gamma_{LV} \left( \frac{dA_{LV}}{dH} - \cos\theta \frac{dA_{SL}}{dH} - A_{SL} \frac{d\cos\theta}{dH} \right) \frac{\sin 2\alpha}{D^2}. \qquad (1)$$

Because of the previously described results $dA_{LV}/dH < 0$, $dA_{SL}/dH > 0$, and $d\cos\theta/dH > 0$, we conclude that the changes of $A_{LV}$ and $\theta$ always contribute to driving the drop movements toward $H$ increasing direction, while the change of $A_{SL}$ can be either a driving or anti-driving factor depending on whether the surface is hydrophilic or hydrophobic. The plots in Figure 4 show examples of respective contributions to the driving force, revealing that the contribution of $\theta$ is negligible on micro- or larger scales (Figure 4(a)) but becomes dominant for smaller $\theta$ on nanoscale (Figure 4(b)). The above described mechanism is uncovered for the first time. Compared with the force arising from the gradient of surface tension, such force arising from the gradient of surface curvature much more likely enlarges with reducing of the cone diameter $D$.

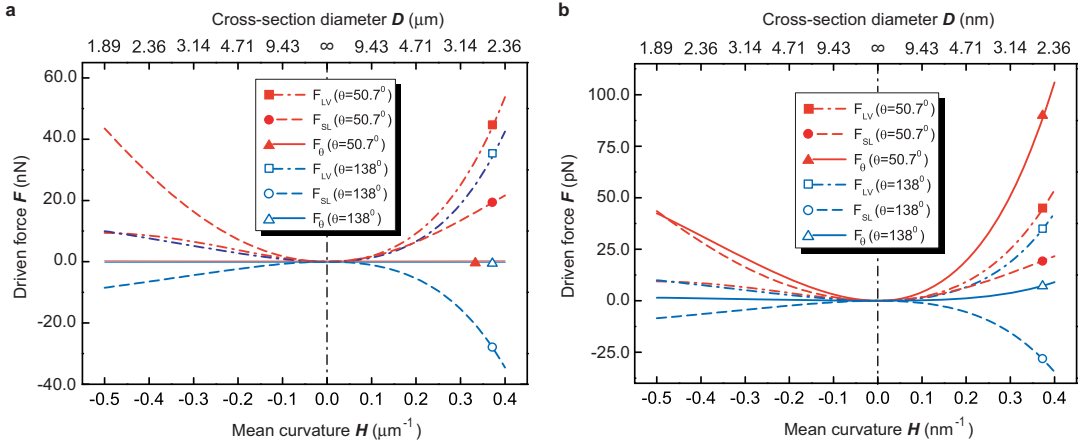

**FIG. 4.** Driving force contributions $F_{LV}$, $F_{SL}$, and $F_\theta$ due to changes of the liquid-vapor interfacial area, solid-liquid interfacial area, and liquid contact anlge caused by gradient of mean curvature. **(a)** A drop with volume $V_0=4\pi r_0^3/3$ and $r_0=1\mu m$ on a microscale cone. **(b)** A drop with volume $V_0=4\pi r_0^3/3$ and $r_0=1$ nm on a nanoscale cone.

Finally, we examine influence of the curvature of a smooth generic solid surface on the liquid adhesive energy or contact angle. At first we analyze the interaction energy, $\Phi$, between a liquid atom (or molecule) near the solid-liquid interface and all the solid surface atoms (or molecules). If we adopt a pair potential (such as van der Waals potential) to model the interaction between the liquid atom and any surface atom at a distance then we can obtain the following approximation (see Supplementary Information for the detailed derivation):

$$\Phi(d) = \frac{\Phi_0(d)}{\sqrt{1-2Hd}}, \qquad (2)$$

where $d$ is the distance of the liquid atom to the solid surface and $\Phi_0$ denotes the total interaction energy as if the surface would be flat. The error ignored from the approximation (2) is of the order $O((\sigma/R_{min})^2)$, where $\sigma$ is the equilibrium distance of the nearest layer of liquid atoms to the solid surface (about 0.32 nm for water and graphene) and $R_{min}$ is the minimum principal curvature radius of the solud surface. Equation (2) reveals that curvature $H$ will significantly affect $\Phi$ if it is on nanoscale. Since the value of $\Phi_0$ is negative, we further see from (2) that enlarging $H$ will always lower down $\Phi$ or, equivalently, increase the adhesive energy. Since the adhesive energy, $\Gamma$, per unit interfacial area can be associated with the contact angle in the form $\Gamma = \gamma_{LV}(1+\cos\theta)$, we see that on nanoscale contact angle $\theta$ is greatly affected by the surface curvature and appears to be a monotonously decreasing function of the single curvature parameter $H$. On the other hand, $\Phi$ is not related with $K$. Now that contribution of the solid-liquid interface to $U$ is the sum of $\Phi$ for all liquid atoms near the interface, Eq. (2) confirms the independence of $U$ upon $K$ for arbitrary smooth generic solid surfaces.

The kind of fast spontaneous movement phenomenon and associated mechanism described here could be useful for rapid cooling, passive water collection, bio-assay, micro chemical sythesis, dose controlled drug releasing, and so on.

**Acknowledgements:** Financial support from the NSFC under grant No.10872114, No.10672089, and No. 10832005 is gratefully acknowledged.

# Supplementary information

For "Ultrafast Drop Movements Arising from Curvature Gradient"

by C.J.Lv, C. Chen, Y.J. Yin, F.G. Tseng, Q.S. Zheng

## S.1 Details of experiments
**Materials:**

Three kinds of glass capillaries with tip/end diameters of 300um/1mm, 300um/1.5mm, and 100um/1mm were used. The first two kinds were specially ordered from a local company (Ching-Fa Company, Taiwan). Only the kind of 100um/1mm was purchased from Sutter Instrument, U.S.A. (the outer/inner diameter is 1mm/0.78mm, and the length is 10cm). The tips of the capillaries were sharpened by a flaming/brown micropipette puller (Model P-1000, Sutter Instrument, U.S.A.) to obtain needle-shape capillaries.

Anhydrous alcohol (99.5%) and anhydrous toluene (99.9%) were purchased from Shimakyu's Pure Chemical (Osaka, Japan) and J. T. Baker (Phillipsburg, NJ, U.S.A). MTS (Methyltrichlorosilane, $CH_3SiCl_3$, 99% purity, catalogue number: M85301) was purchased from Sigma-Aldrich, USA.

**Methods:**
1. *Surface cleaning*

    Piranha solution with a 7:1 volume ratio of $H_2SO_4$ (96%) to $H_2O_2$ (30%) was heated to 90°C and used to preclean surfaces of the glass capillaries. Then dionized (DI) water with a resistance of 20MΩ was used to carefully clean the surfaces.

2. *Surface Modification process*

    Outer surfaces of all the capillaries were pretreated by oxygen plasma at 100 W and 75 mtorr chamber pressure for 300 s. Then the capillaries were put in 0.014 M MTS solution in anhydrous toluene for 75 min under 23°C and 75 % RH environmental conditions. For removing the residual molecules that were not immobilized from the surface, the capillaries were then rinsed in anhydrous toluene, ethanol, a mixed solution of ethanol and DI (1:1), and DI water in a sequence.Compressed dry air was employed to blow dry the capillaries, and finally the capillaries were put into an oven for annealing at 120°C for 10 min [1].

3. *Measurement*

    The static contact angle (CA) of each surface before and after treatment was measured with FTA 200 instrument (First Ten Angstroms, U.S.A.) by using pipette to drop 2 uL DI water on a flat glass slide (Kimble glass, Owen-Illinois, U.S.A.) in

stead of the glass capillaries because of the difficulties in defining contact angle measurement on a curved surface.

Images of the moving 1 μL droplet were captured with a high speed camera (Fastcam-Ultima APX, Photron) coupled with a long focus optical system (Zoom 700X with internal focus vertical illuminator, OPTEM International U.S.A).

**Results & Discussions:**

After piranha solution cleaning, the plain glass capillary surface showed a CA of 28.3°. Then it became highly hydrophilic with a CA of 5° after oxygen plasma treatment.

The capillary surface was modified with a 3D nano-textures of MTS aggregations[1] and treated by oxygen plasma. It finally presented a CA of 0[2]. (as shown in Fig. S1)

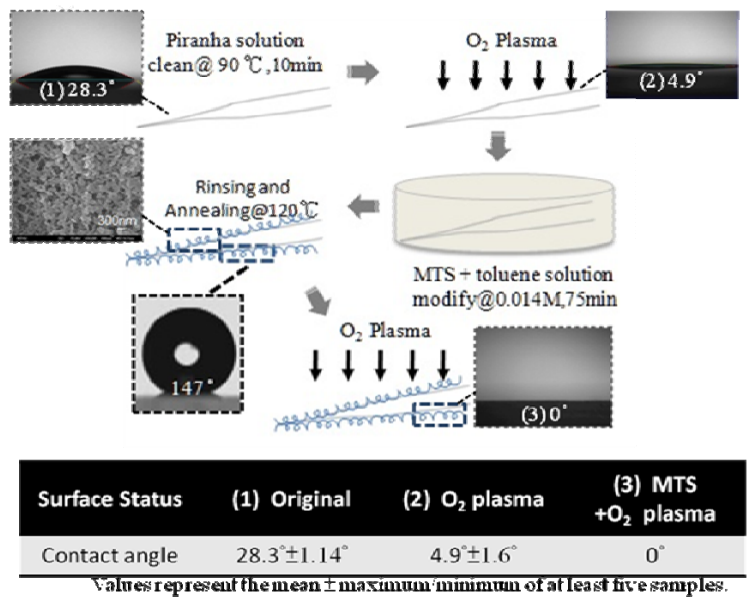

FIG. S1. Capillary surface modification process.

**S.2 Total van der Waals potential energy between a single atom and a general curved surface**

We consider a uniform surface $S$ consisting of mono-layered atoms and a single atom M at distance $d$ from it as illustrated in Fig. S2. $O$ denotes the projection of M on $S$, $S_O$ the tangent plane of $S$ at $O$, and $\{x,y,z\}$ a Cartesian coordinate system [3], of which the origin is located at $O$, $x$- and $y$-axes are tangent to the two principle curvature lines of $S$, and $z$-axis directs to atom $M$.

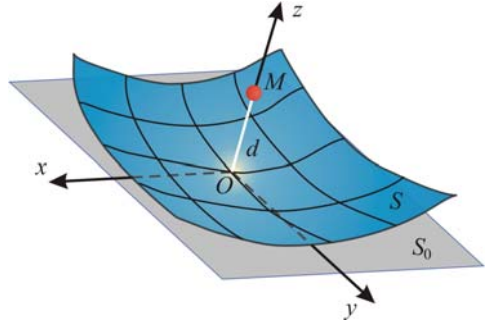

**FIG. S2.** The local cartesian coordinate system {$x,y,z$} to depict the interaction between a particle $M$ and a general curved surface $S$, with $x$-axis and $y$-axis tangential respectively to the principle curvature lines at point $O$, and $z$-axis normal to the surface. $z$-coordinate $d$ of particle $M$ is its distance from $S$.

If we use $z = f(x,y)$ to represent surface $S$, then we have $f(0,0) = f_{,x}(0,0) = f_{,y}(0,0) = f_{,xy}(0,0) = 0$ and

$$f(x,y) \approx \frac{1}{2}(k_1 x^2 + k_2 y^2) \tag{S1}$$

for an area near $O$, where $f_{,x}$ and $f_{,y}$ denote partial derivatives of $f(x,y)$ to $x$ and $y$, respectively, and

$$k_1 = f_{,xx}(0,0), \quad k_2 = f_{,yy}(0,0) \tag{S2}$$

are the two principle curvatures of the surface at $O$. The two principle curvature radii are denoted by $R_1 = 1/|k_1|$ and $R_2 = 1/|k_2|$. And we define $R = \min\{R_1, R_2\}$.

We consider a circular domain $S_\delta$ of plane $S_O$ centered at O with a radius $\delta$. If $\delta$ is of the same order of magnitude as $d$ and $d/R \ll 1$, then distance $r$ between M and any point of $S$ within $S_\delta$ can be approximated, by ignoring $O((d/R)^2)$, as

$$\begin{aligned} r^2 &= x^2 + y^2 + [d - f(x,y)]^2 \\ &\approx x^2 + y^2 + d^2 - 2f(x,y)d, \\ &\approx X^2 + Y^2 + d^2 \end{aligned} \tag{S3}$$

where

$$X = (1 - k_1 d)^{1/2} x, \quad Y = (1 - k_2 d)^{1/2} y. \tag{S4}$$

Area element of $S$ at any point within $S_\delta$ can also be similarly approximated as

$$\begin{aligned} dS &= (1 + f_{,x}^2 + f_{,y}^2)^{1/2} dxdy \\ &\approx dxdy \\ &\approx [(1 - k_1 d)(1 - k_2 d)]^{-1/2} dXdY \\ &= (1 - 2Hd + Kd^2)^{-1/2} dXdY \end{aligned} \tag{S5}$$

where $H = (k_1 + k_2)/2$ and $K = k_1 k_2$ are the mean curvature and Gauss' curvature respectively [3].

If the van der Waals interaction potential[4] between $M$ and the $i$th atom on $S$ at distance $r_i$ is denoted by $\phi(r_i)$, then the total potential is

$$\Phi(d) = \sum_i \phi(r_i) \approx \rho \int_S \phi(r) dS, \tag{S6}$$

where $\rho$ is the density of surface atoms, i.e. the number of surface atoms per unite area, assumed to be constant. By noting that van der Waals interaction is typically negligible as the distance is larger than 1nm, we further have the following approximation

$$\Phi(d) \approx \rho \int_{S_\delta} \phi(r) dxdy \tag{S7}$$

for $d$ in the order of 1nm. Substituting (S5) into (S7) yields

$$\begin{aligned}\Phi(d) &\approx \rho \int_{S_\delta} \phi(\sqrt{X^2 + Y^2 + d^2}) dxdy \\ &\approx \Phi_0(d)(1 - 2Hd + Kd^2)^{-1/2}\end{aligned} \tag{S8}$$

where

$$\Phi_0(d) = \rho \int_{S_O} \phi(\sqrt{X^2 + Y^2 + d^2}) dXdY \tag{S9}$$

is the total van der Waals energy of M when S is flat. Since $Kd^2$ is also of the order $O((\delta/R)^2)$ and negligible, we finally obtain Eq. (2), namely

$$\Phi(d) = \frac{\Phi_0(d)}{\sqrt{1 - 2Hd}}. \tag{S10}$$

**S.3 Supplementary Movies**

**Movie 1, 2, 3** – Experimental observation of the fast spontaneous movements of a 1.0 µl water droplet on conical surfaces

When a 1.0 µl water droplet is released gently enough from the syringe, it will move spontaneously on the conical surface. **Movie 1** (S1.mov, 202KB), **Movie 2** (S2.mov, 117KB), and **Movie 3** (S3.mov, 199KB) represent fast movements of water droplets on Cone I (radius of curvature ranges 0.1-1.0 mm) with three different surface conditions - plain with θ ≈ 28°, $O_2$ plasma treated with θ ≈ 5°, and MTS nanotextured plus $O_2$ plasma treated with θ ≈ 0° - respectively. The actual speeds of the water droplets in experiment is ten times those shown in the videos.

**Movie 4** (S4.mov, 8.21MB) – Molecular dynamics simulation results of water droplet self-motion

The water droplet on the outer/inner conical surface is spontaneously moving toward the larger/smaller cross-section area regardless of the value of the water contact angle ($\theta$=50.7° (hydrophilic), 138° (hydrophobic), or nearly 180° (superhydrophobic)). The droplet is bouncing back when it meets the boundary.